\documentclass{iopconfser}

\usepackage{citesort}
\usepackage{graphicx,amssymb}
\usepackage{hyperref}
\usepackage{graphicx,cite}
\usepackage{changes}
\usepackage{xcolor}
\usepackage{soul}

\begin{document}

\title{Bohmian mechanics: A legitimate hydrodynamic picture for quantum mechanics, and beyond}

\author{\'Angel S. Sanz}

\affil{Department of Optics, Faculty of Physical Sciences, Universidad Complutense de Madrid\\ Pza.\ Ciencias 1, 28040 Madrid, Spain}

\email{a.s.sanz@fis.ucm.es}

\begin{abstract}
Since its inception, Bohmian mechanics has been surrounded by a halo of controversy.
Originally proposed to bypass the limitations imposed by von Neumann’s theorem on the impossibility of hidden-variable models in quantum mechanics, it faced strong opposition from the outset.
Over time, however, its use in tackling specific problems across various branches of physics has led to a gradual shift in attitude, turning the early resistance into a more moderate acceptance.
A plausible explanation for this change may be that, since the late 1990s and early 2000s, Bohmian mechanics has been taking on a more operational and practical role.
The original hidden-variable idea has gradually faded from its framework, giving way to a more pragmatic approach that treats it as a suitable analytical and computational tool.
This discussion explores how and why such a shift in perspective has occurred and, therefore, answers questions such as whether Bohmian mechanics should be considered once and for all a legitimate quantum representation (i.e., worth being taught in elementary quantum mechanics courses) or, by extension, whether these ideas can be transferred to and benefit other fields.
Here, the Schr\"odinger equation and several specific numerical examples are re-examined in the light of a less restrictive view than the standard one usually adopted in quantum mechanics.
\end{abstract}


\section{Introduction}
\label{sec1}

What follows is intended to be a humble tribute to the memory of Basil Hiley, with whom I had the chance to share many good and long conversations over more than a decade about Bohmian mechanics\footnote{
Throughout this work, as I have also done in former works, I will use the denomination ``Bohmian mechanics'', because this is the widespread denomination given to the causal theory developed by Bohm at the beginning of the 1950s (I leave aside here interpretation issues, schools of though, etc., sprung up around this theory).
Nonetheless, it is important to clarify that the idea of using the word ``mechanics'' was, so to speak, uncomfortable to Bohm, as it directly comes into conflict with the very essence of the concept of mechanics in the way it arises and is understood from our everyday intuition.
This is very clearly expressed by Bohm in his renowned book ``Quantum Theory'' \cite{bohm-bk:QTh}, when he talks about the need for a nonmechanical description of quantum systems (see p.~167): ``The entire universe must, on a very accurate level, be regarded as a single indivisible unit in which separate parts appear as idealisations permissible only on a classical level of accuracy of description. This means that the view of the world being analogous to a huge machine, the predominant view from the sixteenth to nineteenth centuries, is now shown to be only approximately correct. The underlying structure of matter, however, is not mechanical.''
To this, Bohm adds: ``This means that the term `quantum mechanic' is very much a misnomer. It should, perhaps, be called `quantum nonmechanics'.''
}, Clifford algebras, the Wigner-Moyal representation, weak values and atom experiments, or David Bohm and the Birkbeck group, but also about everyday life matters.
On the scientific side, which essentially used to turn around Bohm's theory and then divert to may other topics, like those mentioned before (and even to deeper mathematical questions, often hard to be followed but equally fascinating), we frequently ended on converging to the same conclusions: what Bohm proved was not how to demolish the building of quantum mechanics itself, but to stress that this theory can actually be understood within a much wider framework than some of the strict monolithic impositions grounded on the Copenhaguen view.
That is, there is still room in the theory to include many (artificially) banned concepts without abandoning its formal pillars.
Bohm showed that there is a lot of more in there if we pull down unnecessary constraining walls, as time has evidenced, firstly with Bell's inequality and quantum nonlocality, and then through all the related outcomes that we now know as emerging quantum technologies, or the possibility to measure on equal footing in a single experiment both positions (space intensity distributions) and conjugate momenta (transverse momentum distribution) by means of the so-called weak measurements.
In this latter regard, for instance, couldn't have we ever measured positions and momenta at once if we never challenged those unnecessary constraints (i.e., not imposed by the mathematical formalism itself)?

In those conversations with Basil, two recurring questions appeared quite often, which by chance were present in work that each of us has developed over years independently and that seemed to be of common interest to both, namely:
\begin{itemize}
\item Can (what is widely known as) Bohmian mechanics be considered to be a legitimate picture or representation for quantum mechanics?

\item Can we extend its program/philosophy to other fields of physics?
\end{itemize}
These two questions or, to be more precise, trying to find a satisfactory answer to them (not necessarily a definite one) constitutes the driving idea underlying this contribution to honor Basil's efforts to put Bohmian mechanics on the level that it corresponds to within quantum mechanics, beyond ontologies and other metaphysical aspects.
Here, these latter aspects are not going to be considered.
What follows is a fully pragmatic approach to Bohmian mechanics and the consequences it leads to at a formal level, which has nothing to do with hidden variables.
To some extent, paradoxically, this turns out to be a kind of turning back to a Bohr-like conception of the quantum world and its inherent inaccessibility, as it will also be seen, although with a much wider and more flexible conceptual framework than the one originally imposed by the Copenhagen school.

It is not possible to address the above questions disregarding their original source, which traces back, of course, to David Bohm's original 1952 papers\footnote{
It would be unfair neglecting here the role played by de Broglie, whose pilot wave theory preceded Bohm's one by a quarter of a century \cite{debroglie:PhDThesis,deBroglie:JPhysRadium:1927}.
De Broglie settled down the antecedent of suggesting that quantum systems consisted of a wave and a particle, with the wave guiding (``piloting'') the latter's motion and therefore reintroducing the notion of causality in quantum mechanics \cite{broglie-bk:1960}.
However, he did not move forward with the model, which essentially fell into oblivion for about a quarter of a century, after it was fiercely rejected at the Fifth Solvay Conference \cite{5SolvayConf:1928}.
Somehow this parallels Bohm's theory, which was basically forgotten for also about a quarter of a century after being strongly rejected by the quantum community, until the calculations produced by ``the two Chris'' (Chris Dewdney and Chris Phillippidis) showed \cite{dewdney:NuovoCimB:1979,philippidis:NuovoCim:1982} that this theory was beyond speculations.
This pulled down the walls of unfounded criticisms fundamentally grounded on unexplored aspects of the quantum theory and all its unexpected possibilities (indeed, a major push in the development of the current quantum science and technologies!).
Nonetheless, despite the pitiful situation lived for more than fifty years, de Broglie's pilot wave legacy actually transcends its original purpose, as it served to settled the so-called de Broglie's hypothesis that eventually led to associate the wave nature to nonzero mass particles, and motivated the invention of the wave mechanics when Schr\"odinger tried to determine the dynamical equation governing the behavior of de Broglie's matter waves.}
on ``A suggested interpretation of the quantum theory in terms of `hidden' variables'' \cite{bohm:PR:1952-1,bohm:PR:1952-2}.
The basic idea behind this ``suggestion'' was to argue against the objections posed by von Neumann’s theorem on the impossibility of completing or explaining quantum mechanics in terms of hidden variables, at least not without replacing the current theory by another alternative, as it is indicated by von Neumann in \cite{vonNeumann-bk:1932En} (pp.~324-325):
\begin{quotation}
``[\ldots] if there existed other, as yet undiscovered, physical quantities, in addition to those represented by the operators in quantum mechanics, because the relations assumed by quantum mechanics (i.e., I., II.) would have to fail already for the by now known quantities, those that we discussed above.
It is therefore not, as is often assumed, a question of a re-interpretation of quantum mechanics, -- the present system of quantum mechanics would have to be objectively false, in order that another description of the elementary processes than the statistical one be possible.''
\end{quotation}
By simply recasting Schr\"odinger's equation in terms of two real-valued partial differential equations (as Madelung also did half a century earlier \cite{madelung:ZPhys:1926}), Bohm showed evidence of an alternative trajectory-based theory, which did not contradict the formal mathematical structure of quantum mechanics \cite{bohm:PR:1952-1}, but explained the evolution of quantum systems jointly in terms of waves and trajectories, similar to de Broglie's pilot wave theory \cite{broglie:CompRend-1:1926,broglie:CompRend-2:1926,broglie-bk:1960}.
Indeed, in his paper, Bohm claims that:
\begin{quotation}
 ``[\dots], an interpretation of the quantum theory in terms of just such ``hidden'' variables is suggested.
 It is shown that as long as the mathematical theory retains its present general from, this suggested interpretation leads to precisely the same results for all physical processes as does the usual interpretation.
 Nevertheless, the suggested interpretation provides a broader conceptual framework tan the usual interpretation, because it makes possible a precise and continuous description of all processes, even at the quantum level.
 [\ldots], the mere possibility of such an interpretation proves that it is not necessary for us to give up a precise, rational, and objective description of individual systems at a quantum level of accuracy.''
\end{quotation}
Here, three important and deep ideas, crucial in any argumentation in favor or against Bohmian mechanics, are settled:
\begin{itemize}
\item The mathematical structure of Bohmian mechanics is still standard quantum mechanics, so there is no new theory but simply a new perspective to look at quantum systems.

\item There is a broader conceptual framework to describe quantum processes in time and space in a very precise manner, without abandoning the idea of continuity in their evolution (against the Bohr's quantum jumps or von Neumann's collapse), and hence in compliance with the evolution also observed in the Schr\"odinger picture.

\item Because none of the above two facts contradict standard quantum mechanics, there is no reason to renounce to a ``precise, rational, and objective description'' of physical systems at the quantum level.
\end{itemize}

These ideas, which can easily be disclosed from the meaning assigned to the concept of hidden variable, are not only fundamental to the arguments here, but have also been a key issue in the evolution of Bohmian mechanics over the last 75 years, since Bohm's 1952 paper (although the ideas already appear in an incipient manner in Bohm's renowned monograph ``Quantum Theory'' \cite{bohm-bk:QTh}).
It is worth noting, thought, that if we make a simple search in a bibliographical data basis, we observe a curious two-slope trend in the number of citations per year received by Bohm's 1952 papers \cite{sanz:FrontPhys:2019}, with an increase rate of about 1.3~citations per year from the 1960s to the end of the 1990s, to a boost in the rate of about 4.5~citations per year from the 2000s to date.
To understand this shift in trend, we need to split up two aspects here associated with Bohmian mechanics:
\begin{itemize}
 \item Interpretation aspects, which try to use Bohmian mechanics to answer the question of how we face the quantum world at a deep level of understanding (which unavoidably leads to consider the concept of hidden variable) as a ``theory without observers'' \cite{goldstein:phystoday:1,goldstein:phystoday:2}.

 \item Practical applications, which consider Bohmian mechanics as an analysis tool to obtain additional or alternative explanations in different physical contexts.
\end{itemize}
There are different monographs in the literature dealing with the interpretative side of Bohmian mechanics and related ontological questions, with direct applications to a wide variety of problems, from the development of quantum dynamics methodologies to the analysis of relatively complicated physical systems, and also with a combination of both.
As an instance, the interested reader might like to consult Refs.~\cite{bohm-bk:1980,bohm-hiley-bk,holland-bk,cushing-bk:1994,cushing-bk:1996,cushing-butterfield-bk:1999,wyatt-bk,duerr-bk:2009,duerr-bk:2013,sanz-bk-1,sanz-bk-2,oriols-bk,bricmont-bk,norsen-bk}, and references therein.
These two aspects are important to the debate here, because they partially explain the above-mentioned shift in trend.
For a while Bohm's theory stirred a fierce debate on the possibility of hidden variables, which ended with the theory basically forgotten until it was again reconsidered by the Birkbeck group to compute trajectories in many well-known quantum problems.
And it worked very nicely!
So, apart from the hidden-variable debate, Bohmian mechanics proved to be a useful tool to understand many different quantum problems in causal terms without contradicting any of the fundamental principles of quantum mechanics \cite{holland-bk}.
This explains the first trend.
The second arose when the theory became a useful and promising source for trajectory-based methods, on the one hand, mainly led by and within the quantum chemistry community \cite{wyatt-bk} (which arose not directly from Bohm's 1952 paper, but from Madelung's 1926 one and boosted by Wyatt \cite{mccullough:JCP:1969,mccullough:JCP:1971-1,mccullough:JCP:1971-2} and Hirschfelder \cite{hirsch:JCP:1974-1,hirsch:JCP:1974-2,hirsch:JCP:1976-1,hirsch:JCP:1976-2} in the 1970s), and an analysis working tool to understand the physics beyond new quantum problems by numerically simulating their evolution in terms of trajectories \cite{sanz-bk-1,sanz-bk-2,oriols-bk}, leaving completely aside in both cases deeper interpretative aspects.

Although those seem to be very good arguments in favor of a positive answer for the above questions, there are still some thorny questions that should be tackled in connection with Bohm's and Hiley's though.
That is, one would like to find a rationale to some of their ideas, which can be of much interest not only concerning Bohmian mechanics, but that strongly support their formal acceptance.
So, how can we bring to a formal level ideas that seemed to be rather abstract or ontological?
There are three of them that, after reading the interview to Basil by George Musser for Scientific American, ``The Wholeness of Quantum Reality: An Interview with Physicist Basil Hiley'', posted on November 4th 2013\footnote{Find the interview in Scientific American here: {https://www.scientificamerican.com/blog/critical-opalescence/the-wholeness-of-quantum-reality-an-interview-with-physicist-basil-hiley/},\newline or elsewhere here: {https://www.criticalopalescence.com/p/the-wholeness-of-quantum-reality-an-interview-with-physicist-basil-hiley}}, came to my mind and will be considered here to complete the answer to the questions posed above:
\begin{itemize}
 \item Bohm's theory is not the last word. To this, one immediately recalls the causal stochastic approach proposed by Bohm and Vigier in 1954 \cite{bohm:pr:1954}, but also the joint work with Hiley \cite{bohm:PhysRep:1989} or Nelson's stochastic model \cite{nelson:pr:1966}.
 Consequently, this unavoidably leads to reconsider the notion of Bohm's ``particles'' as tracers or mappers of quantum states of motion \cite{sanz:AJP:2012,sanz:JPhysConfSer:2012,sanz:FrontPhys:2019}.

 \item Objects versus processes. This interesting binomial reminds one the dichotomy between two related quantities in quantum mechanics, namely, densities versus quantum fluxes, which, in turn, are directly connected to the way we approach or measure quantum systems, that is, the dichotomy between strong measurements and weak measurements.

 \item Wholeness and undivided universe. These concepts, which led Bohm to use the term quantum ``non-mechanics'' (see footnote~1), are, in my view, directly related to the phenomenon of phase coherence, which not only appears in quantum mechanics, but also in optics.
 In other words, we also see here an instance of how metaphysical notions find a direct connection with important well-defined aspects of quantum mechanics.
\end{itemize}

Below, first by analyzing some formal aspects of quantum mechanics in Sec.~\ref{sec2}, and then re-examining some quantum and non-quantum problems in Secs.~\ref{sec3} and \ref{sec4}, respectively, we will try to find answers to the questions posed above, which, indeed, will help us to provide a well-argued answer to our two original questions.
This is summarized in Sec.~\ref{sec5}.


\section{What Schr\"odinger’s equation ``hides''}
\label{sec2}


\subsection{Rewriting the Schr\"odinger equation}
\label{sec21}

When one deals with Bohmian mechanics, particularly in applications, the two more recurring questions that arise, in one way or another, are what a Bohmian trajectory exactly is and what it is good for.
And, of course, one is always forced to provide a satisfactory reply to something that seems to be evident from the equations such trajectories come from either analytically or numerically.
To address these questions, it is important first to put Bohmian mechanics on the right place.
Leaving aside always debatable interpretative aspects, which would take us off from the solid grounds provided by the mathematical rigor of quantum  mechanics, that place is just that of being another representation or picture of quantum mechanics.
We have quantum states that evolve in time according to Schr\"odinger's wave mechanics, or time-evolving operators following Heisenberg's quantum mechanics.
Originally considered to be antagonist one another, Dirac showed the equivalence between these two representations, which, in turn, led to other representations, like the interaction one.
In closer analogy to classical systems, there are phase-space representations, like the Wigner-Moyal one or Husimi's.
And also trajectory-based representations, like Feynman's path-integral approach, which was also the target of criticisms precisely because of considering the concept of trajectory.
Each one of these representations (and others that we can find when dealing with open quantum systems) provides us with different information about a the quantum system and have also been the source of numerical tool to tackle the study of such systems there where formal analysis is no longer applicable.
As shown by Madelung, what we now widely know as Bohmian mechanics (Bohm's theory, the pilot wave theory, the de Broglie-Bohm theory, the causal interpretation, etc., are all synonyms on a formal basis, that is, leaving aside metaphysical questions) is just another representation to describe quantum systems at the crossroad of Schr\"odinger's wave mechanics, Feynman's path integrals, and the Wigner-Moyal phase-space representation, as it incorporates elements that we can find in all these former approaches, while preserving its own identity and distinction from all of them.

Accordingly, contrary to Bohm \cite{bohm:PR:1952-1,holland-bk}, there is no need to introduce any extra postulate, as its mathematical framework is fully contained in standard quantum mechanics and, moreover, any consequence derived from it, namely, the existence of a velocity field and the related Bohmian trajectories, is consistent and well-defined.
To prove this statement, one simply needs to consider the non-relativistic Schr\"odinger equation for a single system with an associated mass $m$,
\begin{equation}
 i\hbar \frac{\partial \psi({\bf r},t)}{\partial t} = - \frac{\hbar^2}{2m} \nabla^2 \psi({\bf r},t) + V({\bf r},t) \psi({\bf r},t) .
 \label{eq1}
\end{equation}
We talk about a ``system'' and not a ``particle'', because actually this is what Schr\"odinger's equation describes, the time-evolution of a quantum system, regardless of what this system represents.
It can be a particle (strictly speaking, the evolution of its center of mass), but also a set of some relevant degrees of freedom in a particle (e.g., vibrations, rotations, reconfiguration, etc.), while the rest of the particle is totally neglected in the description.

By multiplying Eq.~(\ref{eq1}) on both sides by $\psi^*({\bf r},t)$, and then subtracting to the resulting equation its complex conjugate, we readily obtain the continuity equation that describes the conservation of the probability density:
\begin{equation}
 \frac{\partial \rho ({\bf r},t)}{\partial t} = - \nabla {\bf j}({\bf r},t) .
 \label{eq2}
\end{equation}
This equation indicates that the variations in time undergone by the probability density in a given spatial region correspond to the inside or outside leakage of (quantum) flux across the boundary limiting this region.
In Eq.~(\ref{eq2}), the probability density is
\begin{equation}
 \rho({\bf r},t) \equiv |\psi({\bf r},t)|^2 ,
 \label{eq3}
\end{equation}
in agreement with Born's rule for a quantum system at equilibrium, while the quantum flux is measured in terms of the probability current density \cite{bohm-bk:QTh,schiff-bk}, defined as
\begin{equation}
 {\bf j}({\bf r},t) \equiv \frac{\hbar}{2mi} \left[ \psi^*({\bf r},t) \nabla \psi({\bf r},t) - \psi({\bf r},t) \nabla^* \psi({\bf r},t) \right]
 = \frac{\hbar}{m}\ {\rm Im} \left[ \psi^*({\bf r},t) \nabla \psi({\bf r},t) \right] ,
 \label{eq4}
\end{equation}
which has the dimensions of the probability density multiplied by a velocity.
Indeed, if the right-hand side of Eq.~(\ref{eq4}) is multiplied and divided by Eq.~(\ref{eq3}), we obtain
\begin{equation}
 {\bf j}({\bf r},t) = \frac{\hbar}{2mi} \left[ \frac{\nabla \psi({\bf r},t)}{\psi({\bf r},t)} - \frac{\nabla^* \psi({\bf r},t)}{\psi^*({\bf r},t)} \right] \rho({\bf r},t)
 = \frac{1}{m}\ {\rm Re} \left[ \frac{\hat{\bf p} \psi({\bf r},t)} {\psi({\bf r},t)} \right] \rho({\bf r},t) ,
 \label{eq5}
\end{equation}
where $\hat{\bf p} = - i\hbar \nabla$ is the usual momentum operator.
And momentum divided by mass is just velocity.
Therefore, if we divide both sides of Eq.~(\ref{eq5}) by $\rho({\bf r},t)$, we obtain the new vector field
\begin{equation}
 {\bf v}({\bf r},t) \equiv \frac{{\bf j}({\bf r},t)}{\rho({\bf r},t)}
 = \frac{1}{m}\ {\rm Re} \left[ \frac{\hat{\bf p} \psi({\bf r},t)} {\psi({\bf r},t)} \right] \rho({\bf r},t) ,
 \label{eq6}
\end{equation}
which we denominate the local velocity field, since it measures locally the value of an underlying velocity field that acts on the wave function to cause its diffusion or dispersion throughout the corresponding configuration space.
Or, in other words, using Eq.~(\ref{eq6}), we have
\begin{equation}
 {\bf j}({\bf r},t) = \rho({\bf r},t) {\bf v}({\bf r},t) ,
 \label{eq7}
\end{equation}
which makes more apparent the role of ${\bf v}({\bf r},t)$ as the mechanism responsible for the local dispersion of the probability density at each point ${\bf r}$ at any time $t$.


\subsection{The connection to standard Bohmian mechanics}
\label{sec22}

The next question to address here is where the velocity field (\ref{eq6}) comes from.
So far, we have determined that this field is the source for dispersion in the probability density, and this has been done without abandoning the standard mathematical formalism of quantum mechanics, but only considering well-defined physical quantities, such as $\rho({\bf r},t)$ and ${\bf j}({\bf r},t)$.
There are no extra postulates, approximations, or additional elements that were not present before in the theory.
Hence, in principle, there is no need to introduce additional ``hidden variables''.
If we get back to Bohm's original 1952 paper, the starting point consists of recasting the wave function in polar form \cite{bohm:PR:1952-1,holland-bk}:
\begin{equation}
 \psi({\bf r},t) = A({\bf r},t) e^{i S({\bf r},t)/\hbar} ,
 \label{eq8}
\end{equation}
where both $A({\bf r},t)$ and $S({\bf r},t)$ are real-valued fields.
This nonlinear transformation simply obeys the rule that any complex-valued quantity requires two real-values to be specified (in short, $\{\psi,\psi^* \in \mathbb{C}\} \to \{A,S \in \mathbb{R}\}$).
Substituting the polar ansatz (\ref{eq8}) into the Schr\"odinger equation (\ref{eq1}), one obtains two real-valued differential equations that describe the time-evolution for the new field variables:
\begin{eqnarray}
 \frac{\partial A^2({\bf r},t)}{\partial t} & = & - \nabla \cdot \left[ A^2({\bf r},t)\ \frac{\nabla S({\bf r},t)}{m} \right] ,
 \label{eq9} \\
 \frac{\partial S({\bf r},t)}{\partial t} & = & \frac{\left[ S({\bf r},t) \right]^2}{2m} + V({\bf r}) - \frac{\hbar^2}{2m} \frac{\nabla^2 A({\bf r},t)}{A({\bf r},t)} .
 \label{eq10}
\end{eqnarray}
Equation~(\ref{eq9}) is the usual continuity equation for the probability density if $A^2({\bf r},t) = \rho({\bf r},t)$.
Accordingly, on the right-hand side, we note from Eq.~(\ref{eq7}) that the velocity field can also be written as
\begin{equation}
 {\bf v}({\bf r},t) = \frac{\nabla S({\bf r},t)}{m} .
 \label{eq11}
\end{equation}
We thus see that the source for the velocity field is the local phase developed along time by the wave function (to be distinguished from global phase factors, which are irrelevant concerning the system dynamics) or, to be more precise, its local spatial variations.
Despite this connection, note that we are still moving in the realm of standard quantum mechanics.
The difference with Bohm's original proposal is that Eq.~(\ref{eq11}) is directly postulated \cite{bohm:PR:1952-1,holland-bk}.
In that context, Eq.~(\ref{eq11}) is known as Bohm's guidance condition or Bohm's momentum (it is usually given multiplied by $m$, in the form of a momentum).

Indeed, Bohm's momentum is postulated because of the functional form of Eq.~(\ref{eq10}), which is known as the quantum Hamilton-Jacobi equation given its similarity with its classical counterpart, where $S({\bf r},t)$ would play an analogous role to the mechanical action, $S_{\rm cl}({\bf r},t)$, in classical mechanics.
Consequently, also by analogy with Hamiltonian mechanics, the canonical momentum is simply ${\bf p}({\bf r},t) = \nabla S_{\rm cl}({\bf r},t)$, although we have seen that there is no need to postulate anything.
Bohm's momentum differs from its classical counterpart because of the presence in Eq.~(\ref{eq10}) of an extra term, namely, the third term on the right-hand side, known as Bohm's quantum potential,
\begin{equation}
 Q({\bf r},t) \equiv - \frac{\hbar^2}{2m} \frac{\nabla^2 A({\bf r},t)}{A({\bf r},t)}
  = - \frac{\hbar^2}{4m} \left\{ \frac{\nabla^2 \rho({\bf r},t)}{\rho({\bf r},t)} - \frac{1}{2} \left[ \frac{\nabla \rho({\bf r},t)}{\rho({\bf r},t)} \right]^2 \right\} ,
 \label{eq12}
\end{equation}
the mechanism in ``orthodox'' Bohmian mechanics responsible for the quantumness displayed by quantum systems.
Note that regardless of whether the quantum potential is given with a functional form depending on the amplitude $A({\bf r},t)$ or on the the probability density, which is the truly observable quantity (i.e., experimentally accessible), this potential is a measurement of the curvature displayed by the quantum state itself.
Hence, it provides analogous information to that supplied by the amplitude or the probability density, while the velocity field (or Bohm's momentum) conveys phase-related information, that is, information concerning to space and time phase coherence, which is different to density (occurrence or occupancy) information, although both types of information are tightly connected through the corresponding couplings between Eqs.~(\ref{eq9}) and~(\ref{eq10}).
These couplings establish a mutual feedback between the two fields, while in classical mechanics this coupling is directional: the dynamics of the classical Hamilton-Jacobi equation influences the statistics described by the continuity equation, but the opposite never happens.

If we have a well-defined velocity field, then trajectories can be obtained by direct integration in time.
This is also a formally well-defined and rigorous operation, and hence we are allowed to proceed that way.
Therefore, there is no need at all to postulate the existence of trajectories, because they can always be computed from the equation of motion:
\begin{equation}
 \dot{\bf r}({\bf r},t) = {\bf v}({\bf r},t) = \frac{\nabla S({\bf r},t)}{m} .
 \label{eq13}
\end{equation}
Discrepancies here arise from how the outcomes from Eq.~(\ref{eq13}) must be understood.
In standard Bohmian mechanics, the positions obtained constitute the hidden variables that would serve as a counterexample to von Neumann's theorem.
However, this contradicts the fact that Schr\"odinger's equation is an effective equation that describes only partially quantum system, focusing on those degrees of freedom that keep phase coherence.
In this sense, and given that there is no direct empirical evidence that those positions can be uniquely associated with individual localized system realization, one cannot argue in favor of associating an intrinsic physical reality to such spatial positions.
Indeed, following Bohm himself, these trajectories would represent the averaged or thermalized dynamics of an underlying subquantum stochastic medium out of equilibrium
\cite{bohm:pr:1954,bohm:PhysRep:1989,bohm:PRL:1985}, analogous to some extent to Nelson's quantum stochastic approach \cite{nelson:pr:1966}.

As it has been discussed in earlier works \cite{sanz:AJP:2012,sanz:JPhysConfSer:2012,sanz:FrontPhys:2019,sanz:AnnFondLdB:2021}, without abandoning a realist position, in a broader sense we can still claim that the instantaneous positions described by the Bohmian trajectories provide us with valuable local information about the system dynamics (regardless of what the system itself is describing, namely, translations, vibrations, rotations, dissociation or ionization processes, etc.).
This does not mean that there is, in fact, a true particle or, in general, a ``bit'' of the system traveling along the trajectory, but that the average motion of swarms of non-interacting particles or bits of the system have more chances to accumulate within certain regions than around others.
That is, without revealing the true (empirically inaccessible) motion we get clues of the average dynamics.
The local value of the probability density thus indicates how likely it is that a pathway will be more or less populated, while which average paths are possible is something that determines the local value of the velocity field.
In this hydrodynamic analogy, Bohmian trajectories indicate the positions that some (Bohmian) ``tracers'' follow from a given position at a given time to another position at some other later time.
This picture is thus close to the scenario of classical fluids, where, to understand their dynamics, some tracers are spilt in the fluid we wish to study (think of ink in a fluid flow, for instance).
Perhaps we will never find the means to follow quantum particles, but at least we do count on a reliable method to understand the average pathways that they follow or avoid.

Apart from a single-event description, Bohmian trajectories also allow us to determine which parts of the initial densities end up in which other part of the final densities.
By distributing ensembles of trajectories along close boundaries, we can precisely follow where in configuration space all the probability density enclosed within the boundary will end up at any later time (and also back to the past, due to the reversibility of Schr\"odinger's equation).
This is very convenient in the case of scattering or diffraction problems, as we have a method to associate each final diffraction feature with a region in the initial density distribution \cite{sanz:AnnPhys:2013}.
But it can also be a convenient tool to determine how underlying highly nonlinear or chaotic classical dynamics influence the quantum evolution.
Note that this property is equivalent to what happens in classical mechanics, where the density enclosed in a phase-space region remains constant regardless of the deformations that such a region may undergo over time as a result of the dynamics ruling its evolution.
By setting such boundaries, we are not claiming that all the possible particles enclosed by them will remain inside during the whole evolution, but only that, on average, getting back to Bohm's causal stochastic viewpoint, their number will remain the same, in analogy to the definition of chemical dynamical equilibrium (although the proportion of reactants and products remains stationary over time, the underlying molecular species might be changing from one state to the other, with a net zero balance).


Taking into account all the above, we conclude that Bohmian mechanics constitutes a rather robust picture of quantum mechanics, which does not include any element that was not already present in it, but that, on the contrary, expands the formalism by further developing elements that have always been there.
Within this comprehensive program, there is still an unanswered question to cope with: is the phase field (or its local variations, in the form of a velocity field) a quantity that can be physically measured in the same way that we do with the probability density?
Strictly speaking, there is no phase operator and, in consequence, the phase is not a quantum observable.
At least, this is the widely accepted view in quantum mechanics.
The quantum phase is measured indirectly inferred from the interference traits that we observe in probability densities.
Let us say that this makes a difference between observable quantity and detectable quantity, that is, quantities that do not have an associated operator although still they can be measured or inferred from experimental measures.
As a paradigmatic example, let us consider, for example, the case of the Aharonov-Bohm effect \cite{ehrenberg:PPSB:1949,aharonov:pr:1959,aharonov:pr:1961}, where a shielded magnetic field still has effects on charged quantum particles through the phase-effects induced on their wave function, as Tonomura and coworkers \cite{tonomura:PRL:1986} proved experimentally with electrons in the middle 1980s.
This example is important also to the context of the present work, because it stresses the role of the phase, a quantity without any operator associated, in the dynamics of the electrons without the presence of any direct physical interaction \cite{tonomura:Nature:2008,tonomura:PhysToday:2009}.


\subsection{Observables, detectables and weak values}
\label{sec23}

The above discussion points to an understanding on quantum mechanics where, apart from the well-known quantities determined through the measurement of the expectation values of related operators, we also have additional quantities that are also measurable, in principle, but that are not directly connected to some operators and, therefore, for which we do not have expectation values, strictly speaking.
Still they are there.
Such is the case of the quantum phase, which is typically inferred from the local variations of probability densities, although, as seen above, there are formal tools to specify it as an independent quantity.
Furthermore, we have also seen the strong connection between the dynamics exhibited by quantum systems and the local variations of the phase, which generate (or can be understood as the action of) an underlying velocity field with also local variations.
These two elements, quantum phase and velocity field, are characteristic elements of Bohmian mechanics, as it has been introduced above.
These elements provide us with local information about the quantum system dynamics in a global manner, in the same way that thew wave function also does.
To obtain finer grained details of such dynamics, on the other hand, we can consider the trajectories that arise after integrating in time sets of initial conditions acted by such velocity fields.
These trajectories will supply us with valuable information about the local fluid-type dynamics exhibited by the system as it spreads throughout the corresponding configuration space.
This is essentially the vision of quantum mechanics introduced about a century ago by Madelung \cite{madelung:ZPhys:1926} and rescued by Hirschfelder some fifty years later \cite{hirsch:JCP:1974-1}.

That leads us to an additional question: why are some quantum quantities objectivable in terms of operators while others not, although both arise the same in our experiments?
To answer and understand this question, we need to reconsider the dichotomy between ``object'' and ``process'' posed by Bohm and Hiley.
Specifically, within the present context, the object can be associated with those quantities that are describable in terms of the usual expectation values, i.e., they refer to what we know as quantum observables, according to the postulates of quantum mechanics.
However, in quantum mechanics, as well as in classical mechanics, everything goes from here to here, that is, can be understood as a process.
However, processes, as such have not been related to well-known quantities until the popularization of the so-called weak values, which are the outcome of a so-called weak measurements, that is, a measurement that does not destroy the evolution of the quantum system \cite{aharonov:PRL:1988,aharonov:PRA:1990,sudarshan:PRD:1989,leavens:FoundPhys:2005,wiseman:NewJPhys:2007,hiley:JPhysConfSer:2012}.
To better understand this idea and its implications, consider the measure of the expectation value of a certain operator $\hat{A}$, that is, $\langle \psi | \hat{A} | \psi \rangle$.
If there exists a basis set of eigenfunctions of $\hat{A}$, this expectation value can be written as
\begin{equation}
 \langle \psi | \hat{A} | \psi \rangle = \sum_i \langle \psi | \phi_i \rangle \langle \phi_i | \hat{A} | \psi \rangle = \sum_i \rho_\psi (\phi_i)\ \frac{\langle \phi_i | \hat{A} | \psi \rangle}{\langle \phi_i | \psi \rangle} ,
 \label{eq14}
\end{equation}
where $\rho_\psi (\phi_i) \equiv \langle \phi_i | \psi \rangle \langle \psi | \phi_i \rangle$ describes the spectral decomposition of the system state $|\psi\rangle$ in the basis $|\phi_i\rangle$.
According to Hiley \cite{hiley:JPhysConfSer:2012}:
\begin{quotation}
 ``In recent literature, the quantity $\langle \phi_i | \hat{A} | \psi \rangle / \langle \phi_i | \psi \rangle$ has been called a `weak value' because it magnitude can be found in a so called `weak measurement' if certain criterion are met [1] [11].
 In the context of this paper the concept of `weakness' has no meaning.
 It is merely a transition probability amplitude, [TPA], which, in general, will be a complex number.
 Nonetheless we will, following the convention, continue to call it a weak value.''
\end{quotation}
(This idea is indeed very closely connected to the notion of transition rate introduced by Hirschfelder in the late 1970s \cite{hirschfelder:JCP:1978}.)
Therefore, expectation values, which are obtained from strong of von Neumann measurements, can be recast in terms of a series of weak post-selection measurements in a preferential basis, which do not perturb strongly the quantum system, and still allows us to obtain some extra information about its properties (in such a basis).

Let us consider a basis of eigenfunctions for $\hat{A}$.
After post-selection, we have
\begin{equation}
 \frac{\langle a_i | \hat{A} | \psi \rangle}{\langle a_i | \psi \rangle} = a_i .
 \label{eq15}
\end{equation}
Following the above recipe, Eq.~(\ref{eq14}), we would obtain the usual strong-measurement outcome:
\begin{equation}
 \langle \psi | \hat{A} | \psi \rangle = \sum_i \rho_\psi (a_i) a_i ,
 \label{eq16}
\end{equation}
that is, the average over all possible outcomes $a_i$ properly weighted with the weights $\rho_\psi (a_i)$.
Now, if we consider a basis set that does not consist of eigenfunctions of $\hat{A}$, and then we perform the post-selection in this basis, this leads to
\begin{equation}
 \frac{\langle b_i | \hat{A} | \psi \rangle}{\langle b_i | \psi \rangle}
  = \sum_j \frac{\langle b_i | a_j \rangle \langle a_j | \hat{A} | \psi \rangle}{\langle b_i | \psi \rangle}
  = \sum_j a_j\ \frac{\langle b_i | a_j \rangle \langle a_j | \psi \rangle}{\langle b_i | \psi \rangle} ,
 \label{eq17}
\end{equation}
where $|a_i\rangle$ is an intermediate eigenbasis for $\hat{A}$.
Accordingly, the expectation value of $A$ will be:
\begin{equation}
 \langle \psi | \hat{A} | \psi \rangle = \sum_i \rho_\psi (b_i)\ \frac{\langle b_i | \hat{A} | \psi \rangle}{\langle b_i | \psi \rangle}
 \label{eq18}
\end{equation}
which differs from the functional form (\ref{eq16}) in the fact that now we do not have well-defined outcomes, but rates of change that measure the transition from $|\psi\rangle$ to $|b_i\rangle$ through $\hat{A}$.

This is, precisely, what happens if we consider positions and momenta in configuration space, where the privileged basis is the (continuous) position basis $|x\rangle$.
If we measure the instantaneous expectation value of the position for a system with wave function $|psi(t)\rangle$, the post-selection measurement renders
\begin{equation}
 \frac{\langle x | \hat{X} | \psi (t)\rangle}{\langle x | \psi (t) \rangle}
 = \frac{\int \langle x | \hat{X} | x'\rangle \langle x'| \psi (t)\rangle dx'}{\psi(x,t)}  = \frac{\int x' \delta (x-x') \psi (x',t) dx'}{\psi(x,t)}
 = \frac{x \psi(x,t)}{\psi(x,t)} = x ,
 \label{eq19}
\end{equation}
and the outcome is the well-known expression
\begin{equation}
 \langle \psi (t) | \hat{X} | \psi(t) \rangle = \int \rho_{\psi(t)}(x) (x) dx .
 \label{eq20}
\end{equation}
However, if we wish to measure the expectation value for the position operator $\hat{P}$, we have
\begin{equation}
 \frac{\langle x | \hat{P} | \psi \rangle}{\langle x | \psi \rangle}
  = \frac{\int \langle x | \hat{P} |x'\rangle \langle x' | \psi \rangle}{\langle x | \psi \rangle} dx'
 = \frac{-i\hbar \int \delta (x-x') \nabla_{x'} \psi(x',t)}{\psi(x,t)} dx'
 = \frac{-i\hbar\nabla_x \psi(x,t)}{\psi(x,t)}
 = \frac{\hat{\bf p}\psi(x,t)}{\psi(x,t)} ,
 \label{eq21}
\end{equation}
where $\hat{\bf p} = -i\hbar\nabla$ is the usual momentum operator in the position representation.
Note that, as it is pointed out by Hiley, if $\psi(x,t)$ is recast in polar form, from Eq.~(\ref{eq21}) we readily obtain
\begin{equation}
 \frac{\langle x | \hat{\bf p} | \psi \rangle}{\langle x | \psi \rangle} = \nabla S(x,t) - \frac{i\nabla \rho(x,t)}{2\nabla(x,t)} .
 \label{eq22}
\end{equation}
The first term on the right-hand side is the usual guidance condition (Bohm's momentum), while the second term is the so-called osmotic momentum, which can be related to the underlying dynamics of a sub-quantum stochastic medium \cite{bohm:pr:1954,bohm:PhysRep:1989}, as it was also mentioned by Nelson back to the 1960s \cite{nelson:pr:1966}.


\section{Superpositions, undivided universes, and phase coherence}
\label{sec3}


\subsection{Young's slits and Schr\"odinger cats}
\label{sec31}

The paradigm of the quantum nature is the renowned two-slit experiment, which has in it ``the heart of quantum mechanics'', quoting Feynman \cite{feynman:FLP3:1965}.
As is well known, in this experiment, if there are no markers on either slit indicating which slit the particle crossed, then the latter behaves as a wave.
Experimentally, this means that the distribution of detected particles behind the slits has the form of an interference diagram.
However, as soon as a marker acts on the particles when they cross each slit (for example, assigning them a different polarization state, if we are dealing with photons), such a diagram disappears and we only see the distribution related to a single slit, that is, if there are some short-ranged interaction between the particles and the constituents of the slit support, each slit will produce a nearly Gaussian distribution behind, but if such interaction are negligible we will still observe another wave phenomenon, namely, single-slit diffraction.
Accordingly, we already see that the experiment has nothing of classical in it regardless of external observers or, equivalently, the presence of markers on either slit, as it is commonly stated in undergraduate quantum mechanics courses (although it might resemble it, if single-slit diffraction refers to Gaussian beams \cite{feynman-bk1}).

Nevertheless, this paradigmatic experiment is even trickier than that.
It is also a paradigm for linear superposition, which is a consequence of the linearity in the field (wave function) variable of Schr\"odinger equation.
Accordingly, if $\psi_1({\bf r},t)$ and $\psi{\bf r},t)$ are solutions to this equation, then their linear combination,
\begin{equation}
 \psi({\bf r},t) = \psi_1 ({\bf r},t) + \psi_2 ({\bf r},t) ,
 \label{eq23}
\end{equation}
is also a solution.
As soon as a relative phase between these two partial solutions appears (either from the very beginning or developing with time; either locally, at each point ${\bf r}$, or globally, as a relative phase difference between the two waves), interference appears.
This phase-related phenomenon is detected through the spatial changes induced in the probability density, which reads as:
\begin{eqnarray}
 |\psi({\bf r},t)|^2 & = & |\psi_1({\bf r},t)|^2 + |\psi_2({\bf r},t)|^2 + \psi_1({\bf r},t) \psi_2^*({\bf r},t) + \psi_1^*({\bf r},t) \psi_2({\bf r},t) \nonumber \\
 & = & |\psi_1({\bf r},t)|^2 + |\psi_2({\bf r},t)|^2 + 2 \sqrt{|\psi_1({\bf r},t)| |\psi_2({\bf r},t)|} \cos \varphi({\bf r},t) ,
 \label{eq24}
\end{eqnarray}
where $\varphi({\bf r},t)$ is the above-mentioned relative phase between the two partial waves.
Leaving aside any interpretive consequence, it is clear that the superposition principle is a very efficient mathematical tool, as it allows us to obtain complicated solutions from simpler ones by considering linear combinations of the latter.
This linearity, though, is in striking contrast with the crude reality of the experiment, where there is no possibility to disentangle such partial waves; the quantum system works as an indivisible whole, which involves an inherent strong nonlinear dynamical behavior.
Indeed, what really matters from a experimental point of view is the probability density, not the wave function, because it is the former what is in compliance with the so-called Born's rule.

If we recast each partial wave function in (\ref{eq23}) in polar form (a nonlinear transform itself), as
\begin{equation}
 \psi_i({\bf r},t) = \sqrt{\rho_i({\bf r},t)} e^{iS_i({\bf r},t)/\hbar} ,
 \label{eq25}
\end{equation}
then Eq.~(\ref{eq24}) becomes
\begin{equation}
 \rho({\bf r},t) = \rho_1({\bf r},t) + \rho_2({\bf r},t) + 2 \sqrt{\rho_1({\bf r},t) \rho_2({\bf r},t)} \cos \varphi({\bf r},t) ,
 \label{eq26}
\end{equation}
where
\begin{equation}
 \varphi({\bf r},t) = \frac{S_2({\bf r},t) - S_1({\bf r},t)}{\hbar} .
 \label{eq27}
\end{equation}
On the other hand, the probability current density (\ref{eq4}) acquires the functional form
\begin{eqnarray}
 {\bf j}({\bf r},t) & = & \frac{1}{m} \left\{ \rho_1({\bf r},t) \nabla S_1({\bf r},t) + \rho_2({\bf r},t) \nabla S_2({\bf r},t) + \sqrt{\rho_1({\bf r},t) \rho_2({\bf r},t)}\ \nabla \left[ S_1({\bf r},t) + S_2({\bf r},t) \right] \cos \varphi ({\bf r},t) \right. \nonumber \\
 & & + \left. \hbar \left[ \sqrt{\rho_1({\bf r},t)}\ \nabla \sqrt{\rho_2({\bf r},t)} - \sqrt{\rho_2({\bf r},t)}\ \nabla \sqrt{\rho_1({\bf r},t)} \right] \sin \varphi ({\bf r},t) \right\} .
 \label{eq28}
\end{eqnarray}
Neither the probability density (\ref{eq26}) nor the associated current density are linear superpositions (or can be written in such terms).
This is important, because we are talking about a quantum observable and its associated flux.
If we now try to understand the behavior displayed by these quantities in dynamical terms, by simply dividing the current density by the probability density we obtain the velocity field \cite{sanz:JPA:2008}:
\begin{eqnarray}
 {\bf v}({\bf r},t) & = & \frac{1}{m} \frac{\rho_1({\bf r},t) \nabla S_1({\bf r},t) + \rho_2({\bf r},t) \nabla S_2({\bf r},t) + \sqrt{\rho_1({\bf r},t) \rho_2({\bf r},t)}\ \nabla \left[ S_1({\bf r},t) + S_2({\bf r},t) \right] \cos \varphi ({\bf r},t)}{\rho_1({\bf r},t) + \rho_2({\bf r},t) + 2 \sqrt{\rho_1({\bf r},t) \rho_2({\bf r},t)} \cos \varphi({\bf r},t)} \nonumber \\
 & & + \frac{\hbar}{m} \frac{\left[ \sqrt{\rho_1({\bf r},t)}\ \nabla \sqrt{\rho_2({\bf r},t)} - \sqrt{\rho_2({\bf r},t)}\ \nabla \sqrt{\rho_1({\bf r},t)} \right] \sin \varphi ({\bf r},t)}{\rho_1({\bf r},t) + \rho_2({\bf r},t) + 2 \sqrt{\rho_1({\bf r},t) \rho_2({\bf r},t)} \cos \varphi({\bf r},t)} ,
\end{eqnarray}
which is even more highly nonlinear, not to say the trajectories that emerge after integrating the corresponding equation of motion.

\begin{figure}[t]
 \centering
 \includegraphics[width=14cm]{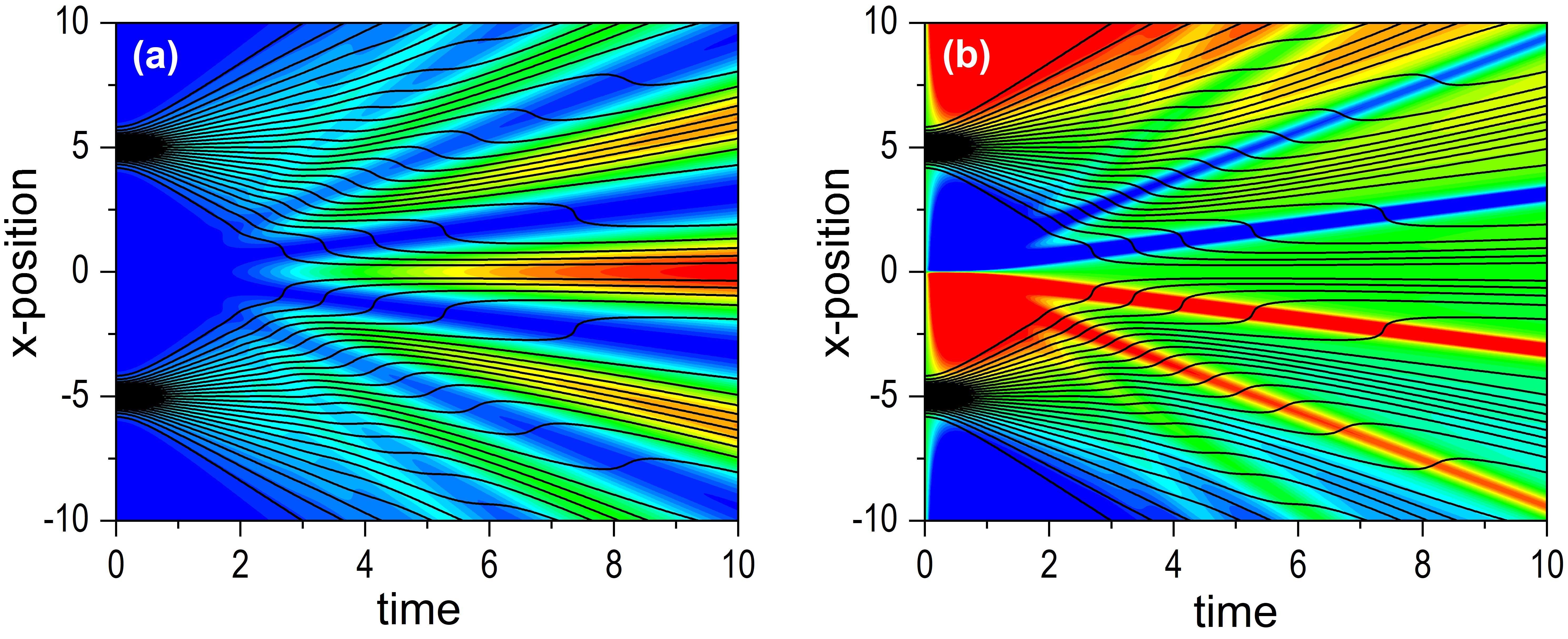}
 \caption{\label{fig1} Probability density (a) and velocity field (b) for a linear superposition of two Gaussian wave packets simulating the transverse diffraction undergone by two Gaussian beams and subsequent appearance of interference traits.
 The Bohmian trajectories associated with the velocity field displayed in panel (a) are superimposed in both graphs to show evidence of the local dynamics induced and, more specifically, how they try to avoid those regions with high values of the velocity and stay in those other with lower values.
 This explains why their statistics results in a series of populated regions surrounded by other regions with negligible density, that is, a series of interference channels that concentrate the average flow of particles.
 In panel (a), the color code associates blue for negligible values of the density and red to higher values; in panel (b); the color code associates blue to negative values of the velocity, red to positive values of the density, and green to values around zero.}
\end{figure}

\begin{figure}[!t]
	\centering
	\includegraphics[width=14cm]{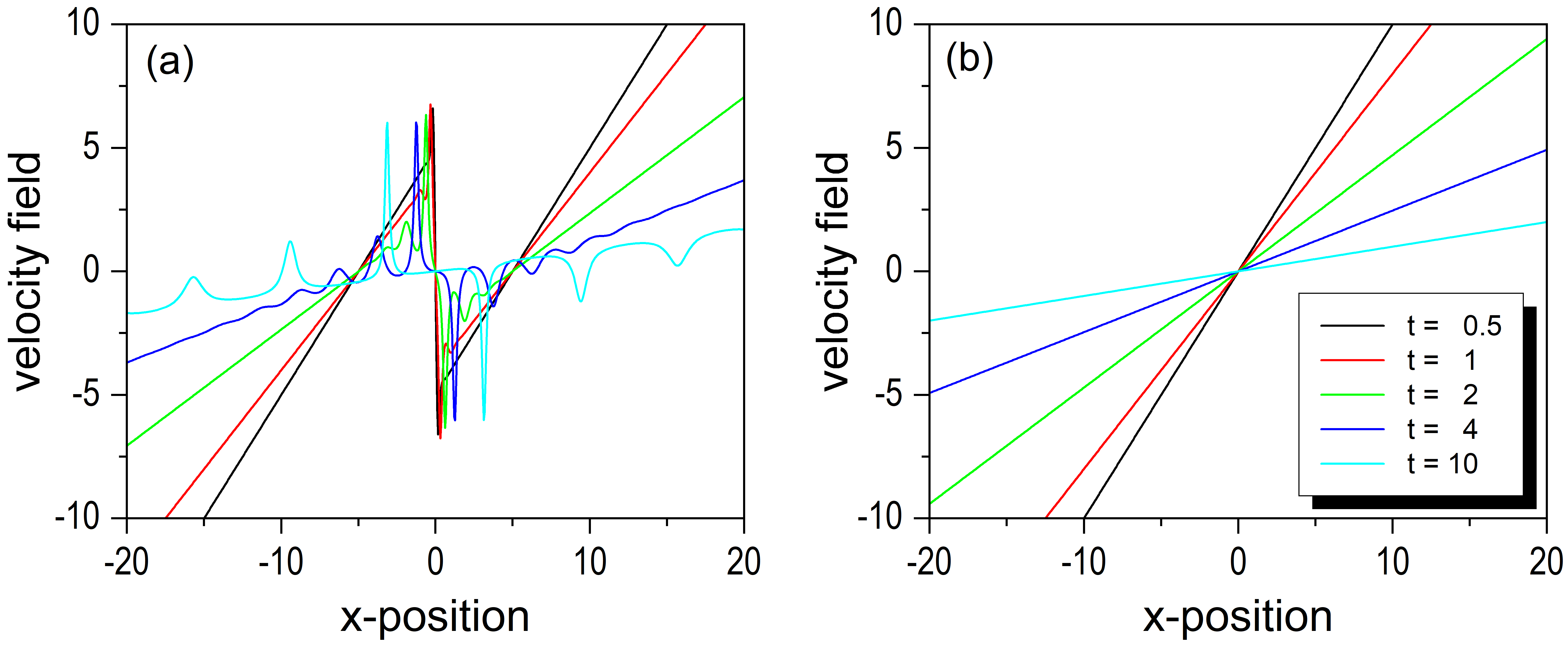}
	\caption{\label{fig2} Transverse cuts of the velocity field at various times for a two-Gaussian wave-packet superposition (a) and a single Gaussian wave packet (b).
		The color code for each transverse cut is indicated in the legend provided in panel (b), which is the same in both cases.
		Although both scenarios exhibit a clockwise rotation of the slope characterizing the velocity field at a given given time, the presence of an intermediate shear at $x=0$ in the two wave-packet superposition is a clear indication that the superposition principle does not hold in the case of the velocity field.
		In this latter case, the system works as a wholeness regardless of whether, mathematically, we describe it in terms of a linear superposition.}
\end{figure}

Figure~\ref{fig1} shows the contour plots of both the probability density (a) and the velocity field (b) in the case of the linear superposition of two Gaussian beams, one of the first problems formerly considered to explicitly illustrate the feasibility of Bohmian mechanics \cite{dewdney:NuovoCimB:1979,philippidis:NuovoCim:1982,holland-bk}.
Leaving aside the question of what the true particle motion is, we readily see that the average or equilibrium dynamics is described by two real fields, namely, a density distribution that accounts for their statistics and a velocity field that, because there are local phase variations, has a major influence on the evolution of the statistical ensemble.
Specifically, in Fig.~\ref{fig1}(a) we observe how, after a while, when the two diffracted Gaussian beams have increased their size and start overlapping, a series of fringes appear.
Now, these fringes would not be there if, at the same that the wave packets spread out, they also develop a phase distribution from nothing (at $t=0$, there is no phase at all).
One could appeal to a quantum potential explanation \cite{sanz:FrontPhys:2019,dewdney:NuovoCimB:1979,holland-bk}, however this sort of ``mechanistic''-type picture involves the probability density again (one hand drawing the other hand, which draws the first one, and so on, like in Escher's famous drawing).
However, the wave function involves two quantities and, therefore, both should be employed to understand this behavior.
This is precisely what the velocity field does, as a direct and evident manifestation of the role played by the quantum phase and, more specifically, the phenomenon of quantum coherence.
In Fig.~\ref{fig1}(b) we observe that, starting from a null value, the development of a phase, a direct consequence of the existence and preservation of spatial and time quantum coherence, provokes a velocity distribution that is not explainable by simply superimposing two waves.
In other words, quantum dynamics transcends the notion of superposition.
Note that, since the very first stages of the system evolution, two distinct regions clearly appear and develop, each one associated with one wave packet, with opposite values for the velocity field.
Only when the two wave packets start overlapping, this constraint relaxes, because a series of regions with strong velocity values appear, which serve as the limit of other regions with lower values of the velocity.
This situation is more clearly seen in Fig.~\ref{fig2}, where several transverse cuts in time of the velocity field are shown.
In panel (a), these cuts show how the velocity field transitions from a shear at the center and a linear trend with high slopes on each side, to series of local maxima and minima aligned along a linear trend with milder slope.
When there is only one wave packet, as it is shown in panel (b), there is only a linear trend that decreases its slope as time increases.
The behavior observed in panel (a) is precisely what Kocsis et all.~\cite{kocsis:Science:2011} observed experimentally fifteen years ago.

If we distribute some initial conditions along the regions covered initially by each wave packet to trace the how the velocity field behaves locally, we readily note that the two swarms spread out quickly (see both panels in Fig.~\ref{fig1}).
Then, because there is an intermediate region with opposite velocities, the trajectories are not allowed to cross the line dividing the figure.
This is indeed the physical origin of the so-called Bohmian non-crossing rule \cite{holland-bk}.
What follows is evident from the topology displayed by the velocity field: in the regions with strong values of the velocity the trajectories will quickly to reach nearby regions with lower velocity values.
Again, this behavior does not come from a linear equation at all.
In this regard, the linearity of Schr\"odinger equation is just an artifact produced by the fact that, instead of being solving a problem with real variables, we are dealing with complex ones \ldots although eventually it is the real quantities what we measure in an experiment.
This is basically what we also do when we consider complex-valued electric and magnetic fields to solve Maxwell's equations; it is a convenient tool, but not the reality.

So, following the above discussion, the conclusion that we extract is that the weirdness of the quantum world is not exactly in the fact that we have wave functions, but on the fields encoded within those wave functions.
Even though if we can only see the shadows cast on the walls of the cave, like the slaves of Plato's cave myth, because there is no way to exactly determine the true motion of quantum particles, still their statistical behavior brings in interesting physics.

Indeed, one wonders what the reality of such velocity fields is.
Or, to be more precise, whether they can experimentally be measured.
By means of weak measurements, as seen above, it seems yes.
In reality (i.e., in the context of real experiments), something of the kind was actually proven by Kocsis and coworkers \cite{kocsis:Science:2011} in the case of light, whose measurements renders something similar to what we find in Fig.~\ref{fig2}.
What this group was able to measure was precisely the profile of the transverse momentum in the two slit experiment with light, which is equivalent to measuring the above velocity field at a series of given times.
Furthermore, even without such an experiment, neglecting the presence of such a field would be equivalent to neglecting the phenomenon of quantum coherence.
Mathematics tell us that coherence allows for superposition (linear combinations); physics tells us that the problem can be subdivided because of the spatial phase coherence sets a relationship since the very beginning in the evolution of quantum systems.


\subsection{Wholeness, phase coherence and bipartite entangled states}
\label{sec32}

\begin{figure}[!t]
	\centering
	\includegraphics[width=\columnwidth]{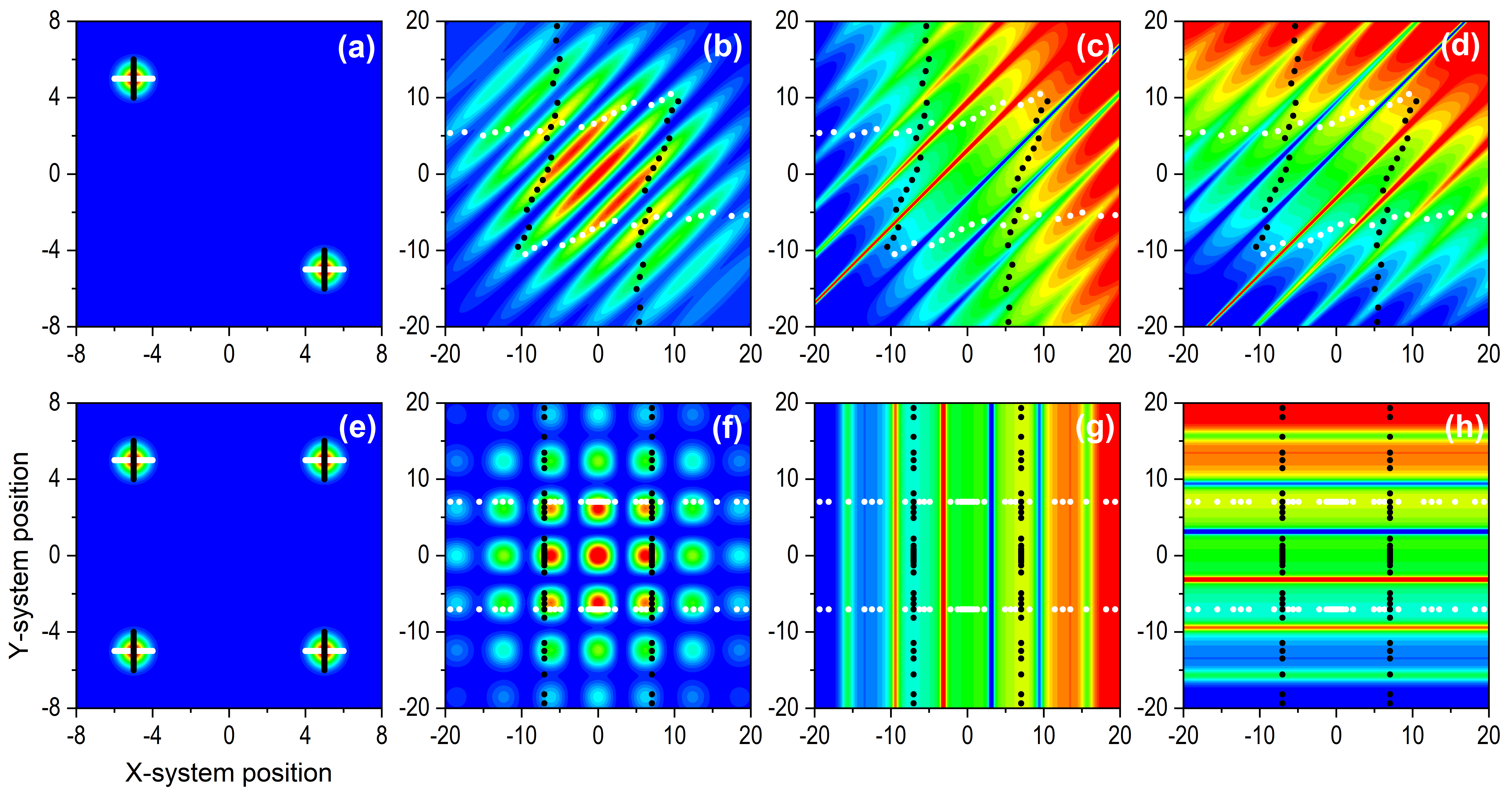}
	\caption{\label{fig3} Propagation of a bipartite continuous entangled state (upper row) and a factorizable two-Gaussian superposition state (lower row).
		The contour plots of the probability density at $t=0$ and $t=10$ are shown in the first and second columns, respectively.
		The velocity field felt by the $X$ and $Y$ systems at $t=10$ are shown in the third and fourth columns, respectively.
		In each plot, the dots represent sets of initial conditions.
		Specifically, the white dots indicate various initial positions for system $X$ keeping fixed the position for system $Y$ at either $A$ or $B$, while dots denote initial positions for system $Y$ keeping fixed, at $A$ or $B$, the position for system $X$.
		In each plot the horizontal axis represents the position coordinates for system $X$, while the vertical axis specifies the position coordinates for system $Y$.
		The color code associated with the contour plots follows the same specifications as in Fig.~\ref{fig1}.}
\end{figure}

We have thus seen that in quantum systems everything acts as a whole and it is difficult to separate each part, because, despite what mathematics tell, this simply means breaking the phase coherence.
This leads us to the question of entanglement, which is often considered the truly distinctive trait of quantum mechanics, as this kind correlation among different quantum systems has no analog in classical physics.
In this case, the same discussion that we have kept above for a single system can also be adopted here, although the idea of phase coherence has to be extended to the full system and not simply to each partner.
To better understand this idea, let us consider the Bell-type state
\begin{equation}
 \psi(x,y) = \frac{1}{\sqrt{2}} \left[ \psi_A(x) \psi_B(y) + \psi_B(x) \psi_A(y) \right] ,
\end{equation}
where $x$ and $y$ make reference to two distinct subsystems, $A$ and $B$ refer to two different locations, and each partial wave is a Gaussian state (further details can be found in \cite{sanz:ENTROPY:2023}).
It is clear that this state cannot be factorized.
At $t=0$, all four Gaussian wave packets are localized around the corresponding sites, $A$ or $B$, as shown in the top row of Fig.~\ref{fig3}(a), and the velocity distribution is zero.
However, as time proceeds, a spatial phase distribution starts developing throughout the full space, which covers and extends beyond the region between the two wave packets.
This leads to a fringe type distribution in the probability density and a velocity distribution that crosses and splits up the space in a series of non-crossing regions perpendicular to the line that joins the two initial distributions.
This is a generalization of the behavior observed for the quantum superposition.
It is this oblique distribution of both densities and velocities what makes impossible the separability of the two subsystems from a dynamical point of view.
As it can be noted in Fig.~\ref{fig3}(b), when we have two factorizable superpositions, there is a symmetry for both the $x$ and the $y$ subsystem that allows such separability, not only concerning the probability density, but also the velocity field.

The consequences of such nonseparability can be better understood, indeed, by studying the behavior of the corresponding Bohmian trajectories.
The dots in Fig.~\ref{fig3} denote sets of initial positions for trajectories along only the $x$ direction or only the $y$ direction.
For a factorizable system, the evolution of one subsystem should have no relevance on the evolution of the other, and vice versa.
This is precisely what we observe in the plots of the lower row of Fig.~\ref{fig2} for the factorizable state, and also in the trajectories displayed for the same system in the lower panels of Fig.~\ref{fig3}.
However, as soon as factorizability disappears, the trajectories start wandering everywhere [but in compliance with the corresponding velocity fields, shown in Fig.~\ref{fig2}(a)] and, therefore, when we project them onto the corresponding subspaces, we immediately notice a lack of correlation between those associated with the wave packet in the site $A$ and those related to the wave packet in $B$, as seen in the upper panels of Fig.~\ref{fig3}.
This does not mean that we the phase coherence has been lost; it has only been lost for a single subsystem, but the coherence for the full system still remains.
What decoherence causes is precisely this loss of phase correlation, because the ``communication'' between trajectories belonging to the same system is dissipated within a bigger configuration space.
It is in this way that, effectively, in general, we are forced to talk about ``undivided universes'', unless we consider a uncorrelated subsystems, which, as we have seen, simply means having lost the phase coherence existing among them or that they never got entangled.

\begin{figure}[t]
 \centering
 \includegraphics[width=12cm]{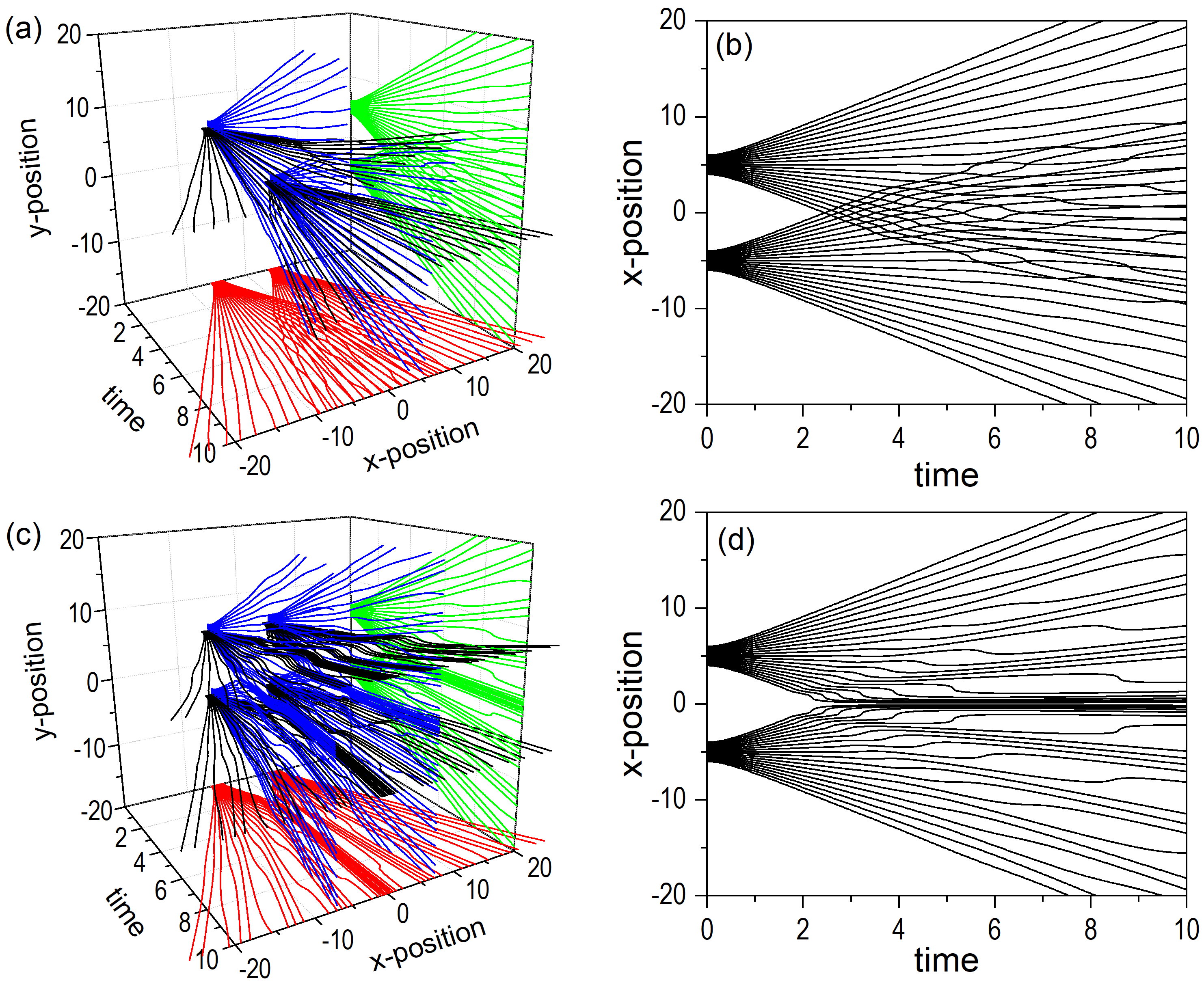}
 \caption{\label{fig4} Bohmian trajectories associated with a Bell-type bipartite continuous variable state (a)-(b), and with a factorizable two-Gaussian superposition state (c)-(d).
 In the left column, the time-evolution of the trajectories in $x$ and $y$ is shown together with the projections onto the $X$- and $Y$-planes, which represent the corresponding subspaces for the $X$ and $Y$ subsystems.
 In the right column, $x$-coordinate of the trajectories for sysbsystem $X$, which represent the projection of this subsystem's trajectories in its corresponding reduced subspace.}
\end{figure}


\section{Bohmian mechanics beyond quantum mechanics}
\label{sec4}


\subsection{Schr\"odinger-like equations in optics}
\label{sec41}

In \cite{feynman:FLP2:1965}, Feynman says:
\begin{quotation}
 ``[\ldots] The equations for many different physical situations have exactly the same appearance.
 Of course, the symbols may be different – one letter is substituted for another –- but the mathematical form of the equations is the same.
 This means that having studied one subject, we immediately have a great deal of direct and precise knowledge about the solutions of the equations of another.''
\end{quotation}
With this in mind, one can consider making applications of Bohmian mechanics to other fields.
After all, if we are able to formulate a problem in terms of a Schr\"odinger-type equation, all the mathematics that lead to the solution are exactly the same.
Another question is how those results are later explained.
And the field of physics where we find a closer analogy to the quantum theory is precisely electromagnetism or, if we take into account the concept of photon, optics.
It has been shown that that, if we start from Maxwell's equations, analogous solutions can be found in terms of the Poynting vector, which involves both the electric and magnetic fields.
For example, in the middle 1970s, Prosser extended this idea \cite{prosser:ijtp:1976-1,prosser:ijtp:1976-2} to the solutions already found by Braunbek and Laukien \cite{braunbek:Optik:1952} to the problem of edge diffraction in terms of the field amplitude and phase.
Later on, solutions in the same terms were also found to the problem of the two-slit interference \cite{sanz:AnnPhysPhoton:2010} or the fringe erasure by polarizing the slits \cite{sanz:JRLR:2010}, for example.

Nevertheless, in those cases where solving the Maxwell equations admits proceeding with the paraxial approximation, equations become isomorphic to those of Bohmian mechanics, since the propagation equation (the so-called Helmholtz paraxial equation) acquires the functional form of a Schr\"odinger equation \cite{sanz:JOSAA:2012,sanz:ApplSci:2020,sanz:JPhysConfSer:2024}:
\begin{equation}
 i\frac{\partial \psi({\bf r}_\perp,z)}{\partial z} = - \frac{1}{2k} \nabla_\perp^2 \psi({\bf r}_\perp,z) ,
 \label{eq33}
\end{equation}
where time is replaced by the longitudinal coordinate and the Laplacian by the transverse Laplacian,
\begin{equation}
 \nabla_\perp^2 \equiv \frac{\partial^2}{\partial x^2} + \frac{\partial^2}{\partial y^2} .
\end{equation}
Physically, in this reparameterization the longitudinal coordinate ($z$, in this case) denotes the direction along which most of the electromagnetic energy (photons) flow, while the dispersion of the wave field takes place within the transverse plane, that is, angular effects due to an effective transfer of energy between the longitudinal and transverse coordinates are neglected.
Also note that, in the case of nonzero mass systems, the relationship between the time and the longitudinal coordinate is given by the de Broglie relation \cite{sanz:AOP:2015}, $p = \hbar k = h/\lambda_{\rm dB}$, where $\lambda_{\rm dB}$ is the de Broglie wavelength:
\begin{equation}
 z = \frac{\hbar k}{m}\ t = \frac{h}{m\lambda_{\rm dB}}\ t ,
\end{equation}
which readily leads us to the usual Schr\"odinger equation (in two coordinates).
In optics, however, the second equality is recast as
\begin{equation}
 z = \frac{\hbar k}{m}\ t = \frac{h n}{m\lambda_0}\ t ,
\end{equation}
where $n$ is the refractive index of the medium and $\lambda_0$ is the light wavelength in vacuum.
This expression is actually quite convenient, for example, in potential-type modeling of optical fibers \cite{sanz:JOSAA:2012}.

This formulations thus allows us to obtain the corresponding electromagnetic flow trajectories in terms of the longitudinal coordinate by integrating the effective transverse velocity field [Sanz, JPhysConfSer 2024]
\begin{equation}
 \frac{d\psi({\rm r}_\perp,z)}{dz} = {\bf v}({\bf r}_\perp,z) = \frac{1}{k} {\rm Re} \left[ \frac{-i\nabla_\perp \psi({\bf r}_\perp,z)}{\psi({\bf r}_\perp,z)} \right] ,
\end{equation}
which presents the same functional form of the usual Bohmian velocity field for nonzero mass quantum systems.
Note that here times does not appear because of the highly fluctuating fields involved in problems with light.
This is, indeed, the correct expression to obtain the trajectories in experiments like the one reported in \cite{kocsis:Science:2011}, where this extension of Bohmian mechanics to paraxial optics was considered.

Taking this into account, in general, we can thus set a general theoretical framework to deal with either quantum systems or paraxial problems in optics \cite{sanz:JPhysConfSer:2024}, where the corresponding scalar wave field would be described by the Schr\"odinger-type equation:
\begin{equation}
 i\frac{\partial \psi(\rho,\xi)}{\partial \xi} = - \frac{1}{2} \nabla_\rho^2 \psi(\rho,\xi) ,
\end{equation}
and the velocity field by
\begin{equation}
 \frac{d\psi(\rho,\xi)}{dz} = {\bf v}(\rho,\xi) = \frac{1}{k} {\rm Re} \left[ \frac{-i\nabla_\rho \psi(\rho,\xi)}{\psi(\rho,\xi)} \right] .
\end{equation}
Integrating this field along the parameterizing coordinate $\xi$ ($t$ or $z$) for specific initial conditions one obtains corresponding trajectories $\rho(\xi)$.


\subsection{Airy-beam propagation}
\label{sec42}

As an example of application of these ideas, let us consider the solutions to the free-propagation equation with initial condition given by
\begin{equation}
 \psi(x,0) = Ai(x)
\end{equation}
where $Ai(x)$ is an Airy function.
As it was shown by Berry and Balazs by the end of the 1970s \cite{berry:AJP:1979}, the solutions are nondispersive wave packets that retain the shape of the original Airy function, which displays an accelerating motion along the $x$-coordinate (i.e., every part of the wave packet feels a displacement proportional to $t^2$).
These type of wave packets were first observed with light, in 2007, by Christodoulides and coworkers \cite{christodoulides:PRL:2007,christodoulides:OptLett:2007}, although replacing time by the longitudinal coordinate $z$.
These solutions have the form
\begin{equation}
 \psi(x,z) = e^{i(x-z^2/6)z/2} Ai(x-z^2/4) ,
\end{equation}
where we can see that, effectively, the argument of the amplitude shows a quadratic dependence on $z$, making the wave packet to move forward along the transverse direction according to $z^2$.
It is easily seen that, since the local phase variations depend bilinearly on both coordinates, the effective velocity field is going to be linear with the longitudinal coordinate:
\begin{equation}
 v(x,z) = \frac{dx}{dz} = \frac{z}{2} .
\end{equation}
This effective equation of motion has simple analytical solutions, namely, parabolic trajectories \cite{sanz:JOSAA:2022},
\begin{equation}
 x(z) = x(0) + \frac{z^2}{4} ,
\end{equation}
which explain very nicely and at a local level how and why these wave packets propagate in the way the do it.

\begin{figure}[t]
 \centering
 \includegraphics[width=14cm]{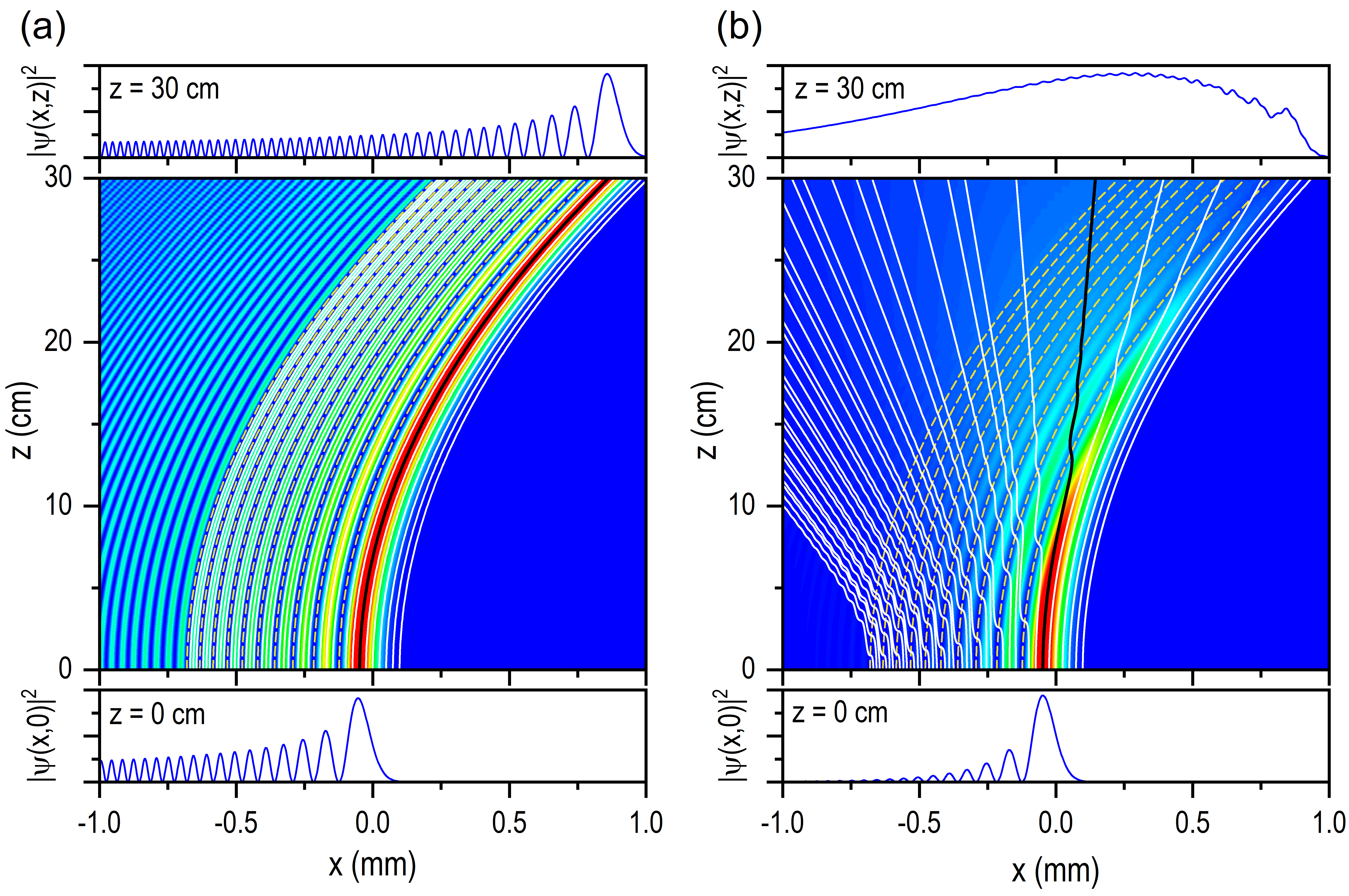}
 \caption{\label{fig5} Propagation of an ideal Airby beam (a) and a finite-energy Airy beam (b).
 The contour plots represent how the transverse intensity distribution varies along the longitudinal coordinate, both given in physical units (i.e., following data from the experiment \cite{sanz:JOSAA:2022}).
 Superimposed to the contour plots, solid lines represent Bohmian trajectories illustrating the beam propagation.
 Dashed lines, on the other hand, indicate the position of the nodal point as a function of $z$ in the ideal case (a), which serve to better understand the kinks observed in the trajectories associated with the finite-energy Airy beam (b).
 These trajectories have also been computed using Eq.~(\ref{eq45}), although it is clear that, in regions where the intensity vanishes, there cannot be trajectories.
 Finally, the black solid line represents the trajectory that starts with initial condition at the position where the initial beam reaches its main intensity maximum.
 Concerning the plots below and on top each contour plot, they represent, respectively, the initial intensity distribution and the final one at $z=30$~cm.}
\end{figure}

Furthermore, this trajectory-based description contributes to settle some objective tools to study the propagation of these wave packets, not only in the case of ideal cases, but more importantly in those cases affected by modifications, where the phase variations and their effects on the subsequent propagation become capital.
After all, in optics phase coherence is even more important than amplitude distributions, particularly in the design and generation of structured light beams.
So, consider now the case of a damped Airy wave packet at $z=0$, i.e., an Airy function with a tail that decreases exponentially fast backwards:
\begin{equation}
 \psi(x,0) = Ai(x) e^{\gamma x} .
\end{equation}
The analytical propagated solution is given by the wave field \cite{christodoulides:OptLett:2007}:
\begin{equation}
 \psi(x,z) = e^{i(x-z^2/6)z/2 + \gamma(y - i\gamma z/2)} Ai(y) ,
\end{equation}
where
\begin{equation}
 y \equiv x - \frac{z^2}{4} + i\gamma z
\end{equation}
is now a complex valued variable.
Obviously, if the Airy function has now a complex argument, the velocity field is also going to depend on this argument, and not only of the phase prefactor, as before.
Specifically, now the effective velocity field reads as
\begin{equation}
 v(\tilde{x},\tilde{z}) = \frac{d\tilde{x}}{d\tilde{z}} = \frac{\tilde{z}}{2} + \frac{\partial}{\partial \tilde{x}}\ {\rm arg} \left[ Ai(\tilde{y}) \right] ,
 \label{eq45}
\end{equation}
which cannot be integrated analytically.
Still we obtain valuable information from it, because it tells us that, whenever there is any minor perturbation on the original Airy beam, it translates into a phase effect that can be directly detected by means of either the additional terms that appear in the velocity field or by the disturbances from parabolicity displayed by the integrated trajectories.
In Fig.~\ref{fig5} the propagation of both an ideal Airy beam (a) and a truncated Airy beam (b) are shown.
As it can readily be noticed, while the ideal beam follows a perfect parabolic lateral displacement along the $x$-direction as the energy propagates forward along the longitudinal coordinate $z$, in the case of the truncated beam the rearmost trajectories start moving backwards so that, at long distances from the starting point, the shape of the beam has completely been lost and the distribution looks more like that of a Gaussian.
This effect, as it is explained in \cite{sanz:PRA:2022}, has to do with the fact that the pressure (just using the same concepts introduced by Takabayashi in the 1950s \cite{holland-bk,takabayasi:ProgTheorPhys:1952}) arising from the back part of the wave packet has decreased enough or even disappear because of the exponential truncation, and this relaxation allows forward contributions for finding a way to propagate backwards.
In other words, an ideal Airy beam keeps its form because of the huge push felt by the infinite back tail, which translates into a quadratic lateral displacement and the conservation of the overall shape, all of this being objectively evidenced by the parabolicity of the corresponding trajectories.

\begin{figure}[t]
 \centering
 \includegraphics[width=\columnwidth]{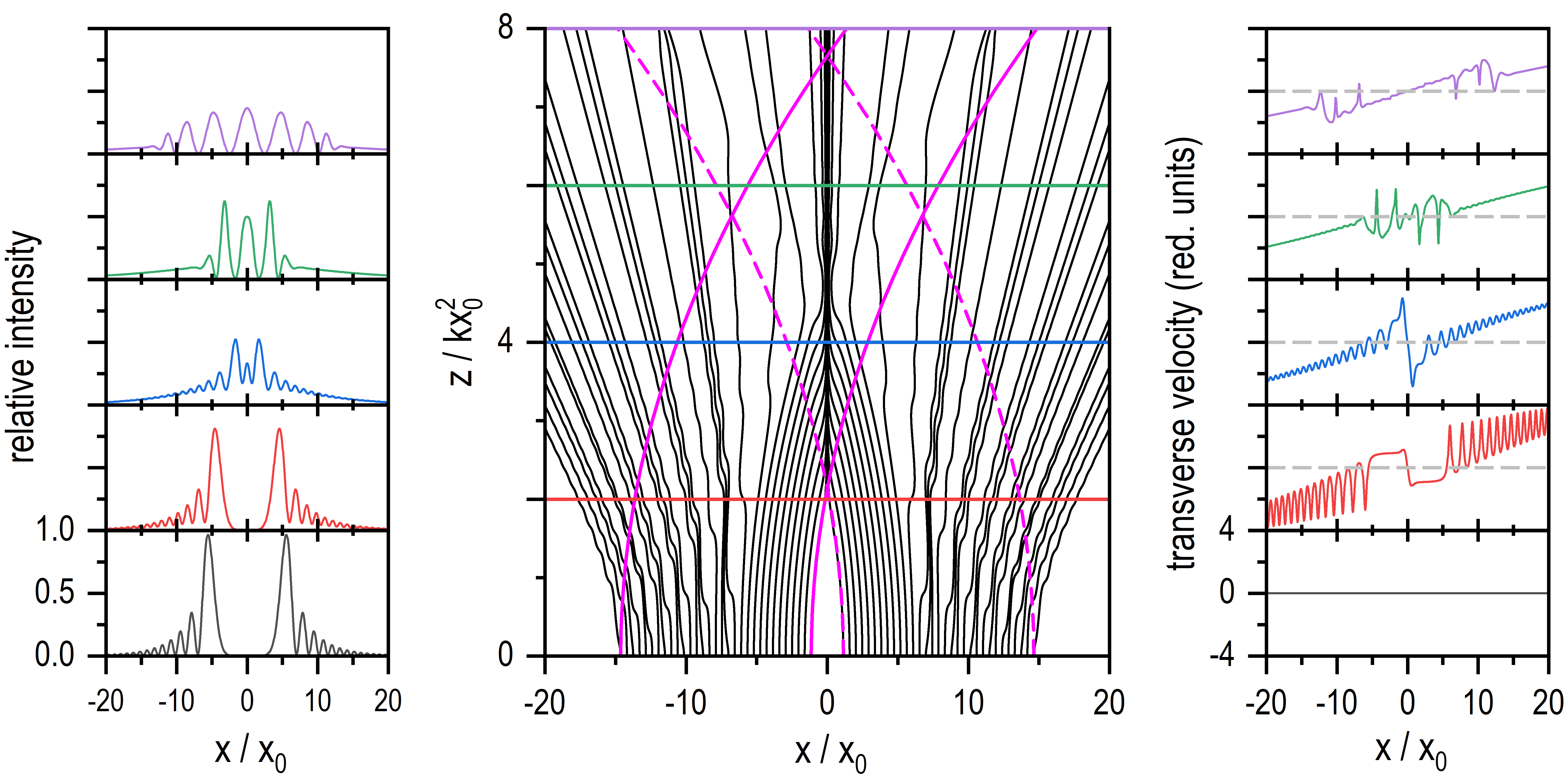}
 \caption{\label{fig6} In the central part, Bohmian trajectories associated with a superposition of two counter-propagating finite-energy partially-coherent Airy beams.
 The magenta solid lines indicate the boundaries within which the trajectories should be confined in the ideal case (the magenta dashed lines denote the same, but for the right-side beam).
 The panels to the left and to the right show, respectively, the transverse intensity distribution (in relative terms, i.e., with respect to the maximum value at $z=0$) and the effective transverse velocity field at various values of $z$.
 These values of $z$ are indicated with the colored horizontal lines in the central panel.
 For simplicity, both coordinates are given in reduced units, $x$ as divided by a typical transverse length value $x_0$ and $z$ by $kx_0^2 = 2\pi n x_0^2/\lambda_0$ (see text for details).}
\end{figure}

The analysis can be further extended to include cases of structured light beam interference \cite{sanz:JOSAA:2022} or even partially incoherent structured light \cite{sanz:OLT:2025}, as it shown in Fig.~\ref{fig6}, where the evolution of two counter-propagating finite-energy and partially coherent Airy beams is shown \cite{sanz:OLT:2025}.
This thus probes to be a very convenient and powerful analysis tool in optics, quantum optics, and photonics, where actually it is possible to act directly on the system phase by means of spatial light modulators experimentally.
In other words, this is indeed the ideal workbench to make real the ideas involved in Bohmian mechanics.

\section{Concluding remarks}
\label{sec5}

The discussion in this work has pivoted around two key ideas, namely, whether Bohmian mechanics is a quantum picture on equal footing as other pictures (Schr\"odinger, Heisenberg, Dirac, Wigner-Moyal, etc.) and also whether its program can be transferred to other fields of physics.
These questions have been addressed always keeping in mind three key aspects, namely:
\begin{itemize}
 \item Bohm’s theory is not the last word.
  We consider that Bohmian mechanics is a picture or representation that does not describe the true motion of real particles, as this is not empirically provable.
  Rather, it provides us with the tools necessary to explain and understand the evolution of quantum systems in statistical terms, analogous to those that we also find in our description of fluids, where the purpose is not accounting for the detailed analysis of a single fluid component, but as a whole.
  This vision is perfectly compatible even with the standard conception of quantum mechanical processes, although it offers us an alternative perspective, which allows us to follow pieces of the quantum ``fluid'' in a very precise and unique manner.

 \item Objects versus processes.
 Bohmian mechanics allows us to go beyond the usual quantum observables and to investigate the evolution of quantum systems in terms of structures that develop in a configuration space in a continuous manner.
 The emergence of these structures or fields (densities and velocities) becomes more accessible and rational by using trajectories (Bohmian ones) that provide us valuable dynamical information about their origin and continuous change in time.
 Experimentally, the so-called weak measurements make accessible to us these structures, since their ``weakness'' does not cause a sudden break in the coherent evolution of the system, as a von Neumann measure does, but simply determining a property related to the change of the system.

 \item Wholeness and undivided universe.
 Bohmian mechanics unveils the leading role of the quantum phase in the dynamical behavior exhibited by quantum systems; the probability density or any other quantum observable are just a consequence of how such phase develop over time.
 In this regard, we cannot put on the same level the mathematical notion of superposition (a suitable working tool) and the physical concept of phase coherence (the truly distinctive trait of any quantum system, regardless of whether it involves one single particle or degree of freedom, or many of any of them).
 Phase coherence generates out of phaseless systems from a matching (phase) correlation condition that has to be satisfied at every spatial point and at each time, which is implicit in Schr\"odinger's equation in the way that amplitudes and phases couple [more evident in the case of the mutual coupling between Eqs.~(\ref{eq9}) and (\ref{eq10})].
\end{itemize}

Therefore, leaving aside philosophical implications or conceptions concerning the reality of Bohmian trajectories, we are unavoidably facing a robust quantum approach that cannot be disentangled from standard quantum mechanics.
Bohmian mechanics is quantum mechanics, plain and simple.
It only provides us with a different alternative perspective of quantum phenomena, of course, but there is nothing in it of controversial or illegitimate, because its trajectories do not contravene any essential principle or postulate of quantum mechanics.
On the contrary, it contributes to increase the scope of quantum mechanics by pulling down artificial bounds, as Feynman's paths also did almost eighty years ago.

Now, given the capital role played by the phase in the problems dealt with quantum mechanics it is clear that the range of applicability of Bohmian ideas will also transcend the quantum realm and will reach those areas where phase also plays a major role.
This is the case of electromagnetism and, in particular, optics.
Indeed, in this other realm, usually regarded as ``classical'', not only we find descriptions that are totally isomorphic to using Schr\"odinger's equation, but we can also manipulate light through its phase, which makes even more interesting the application of the Bohmian program.
And, conversely, this program proves to be very beneficial in the study of light beam design and propagation.


\section*{Acknowledgments}

This contribution, based on the talk of the same title given at University College London on July 1st 2025, constitutes a late tribute to the memory and friendship of Basil Hiley, whom I had the chance to learn a lot through the multiple conversations we had in the last fifteen years.

Financial support from the Spanish Agencia Estatal de Investigaci\'on (AEI) and the European Regional Development Fund (ERDF) (Grant No. PID2021-127781NB-I00) is acknowledged.




\providecommand{\newblock}{}

\end{document}